\documentclass[pre,preprint,showpacs]{revtex4-1}

\usepackage{graphicx}
\usepackage{amsmath,amssymb}
\usepackage{bm}
\usepackage{cancel}
\usepackage{color}

\usepackage{hyperref}
\usepackage{cleveref}

\usepackage{subfigure}
\usepackage{placeins}

\begin{document}
 
\title{STM driven transition from rippled to buckled graphene in a spin-membrane model}
\author{M Ruiz-Garc\'{\i}a$^1$, L L Bonilla$^1$ and A Prados$^2$}

\affiliation {$^1$G. Mill\'an Institute, Fluid Dynamics, Nanoscience and Industrial
Mathematics, Universidad Carlos III de Madrid, 28911 Legan\'es, Spain}
\affiliation{$^2$ F\'{\i}sica Te\'{o}rica, Universidad de Sevilla,
Apartado de Correos 1065, E-41080, Sevilla, Spain}

\date{\today}
\begin{abstract}
  We consider a simple spin-membrane model for rippling in
  graphene. The model exhibits transitions from a flat but rippled
  membrane to a buckled one. At high temperature the transition is
  second order but it is first order at low temperature for
  appropriate strength of the spin-spin coupling. Driving the system
  across the first order phase transition in nonequilibrium conditions
  that mimic interaction of the graphene membrane with a STM tip
  explains recent experiments. In particular, we observe a reversible
  behavior for small values of the STM current and an irreversible
  transition from flat rippled membrane to rigid buckled membrane when
  the current surpasses a critical value.  This work opens the
  possibility to test mechanical properties of graphene under
  different temperature and electrostatic conditions.

\end{abstract}





\maketitle

\maketitle

\section{Introduction}

Rippling in suspended graphene \cite{mey07} is one of its most
compelling mechanical properties, and is usually linked to the
impossibility of finding a perfect crystal in two dimensions
\cite{mermin}.  Thus, the out-of-plane displacements would make it
possible to stabilize the graphene sample. The understanding of this
rippling has triggered a great amount of theoretical work, both
starting from first principles
\cite{fas07,abe07,kim08,gazit09,san-jose,gui14,gonzalez,ByR16} and
using simple statistical mechanics models
\cite{jsm12bon,pre12bon,Ruiz-Garcia,prb12bon,schoelz15}.

The typical length of these graphene ripples, which do not have a
preferred direction \cite{mey07,pss09ban}, is in the nanometer
range. Moreover, they modify the electronic band structure of graphene
\cite{PRB08gui} and are expected to have a prominent role in its
electronic transport \cite{PTRS08kat}. There have been many attempts
to characterize ripples as equilibrium phenomena, connecting them with
thermal fluctuations \cite{fas07,abe07} and the electron-phonon
coupling \cite{kim08,gazit09}. Also, some authors have tried to
describe their curvature starting from first principles
\cite{gonzalez,gui14}.

Recently, there has been a growing interest in buckling of suspended
graphene both for theoretical reasons and for its role in designing
graphene-based devices.  There are many experimental studies of
buckled graphene sheets
\cite{SN88,CN93,leh13,kot14,rob14,BCGW15,lin12,svensson11,xu14,schoelz15},
\textrm{including some very recent ones in which  Molecular
  Dynamics (MD) simulations are also carried out
  \cite{neek-amal,ackerman}. Buckling} can be produced by the application of strong enough
electrostatic forces, as in refs.~\cite{lin12,svensson11}, \textrm{by}
the combination of heating and an electrostatic force, as in
ref.~\cite{schoelz15} \textrm{or even by only heating the sample, as
  in the ``mirror'' buckling observed in \cite{ackerman} by means of
  MD simulations.}

\begin{figure}
\centering
\includegraphics[width=4in]{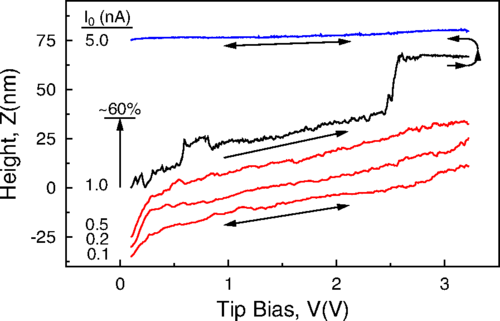}
\caption{Height of the STM tip on the graphene sheet versus applied
  voltage for different values of the tunneling
  current. Curves are slightly offset from each other for
    clarity. From ref. ~\cite{schoelz15}.}
\label{schoelz}
\end{figure}

Buckling upon heating a graphene sample has been systematically
investigated in ref.~\cite{schoelz15} by using scanning tunneling
microscopy (STM).  Specifically, the tip of the microscope is centered
on a suspended sample that is initially flat on average although it is
surely covered with ripples \cite{mey07}. Application of a voltage
bias $V$ between the STM tip and the membrane has a twofold effect:
(i) it induces a tunneling current that locally heats the sample, and
(ii) it produces an electrostatic interaction between the tip and the
sample. Experiments show that the suspended graphene sheet experiences
a transition from ``floppy'' rippled-flat to ``rigid'' buckled
state. The membrane height $Z$ is plotted in Fig.~\ref{schoelz} as a function of the voltage
bias $V$ for several values of the tunneling intensity $I$.  On the one hand, for ``small'' values of $I$, the
height $Z=Z(V)$ is a monotone increasing and continuous function of
$V$. The membrane is rippled and its behavior is reversible: the same
curve $Z(V)$ is observed whether the voltage bias increases from $0$
to a certain value $V_{\max}$ or decreases from $V_{\max}$ to $0$. On
the other hand, once the current is kept constant at a high enough
value, increasing the bias causes the sample to buckle irreversibly:
once a sufficiently large value $V_{\max}$ is reached, the sample
remains buckled as the bias is decreased back from $V_{\max}$ to zero.

Schoelz {\em et al} proposed a phenomenological Ising model to explain
their experimental results~\cite{schoelz15}.  In their model, each
local spin $\sigma_{ij}$ represents one ripple composed of $\sim 1000$
carbon atoms and the value of the spin indicates the curvature of the
ripple. The energy of this Ising system has two
contributions. Firstly, a nearest neighbor spin-spin interaction, with
a coupling constant $J$ that depends on the total magnetization
$M=\sum_{ij}\sigma_{ij}$.  The second contribution to the energy is an
interaction of the spins with an external field $h=h_0 e^{-r/\xi}$,
where $h_{0}$ is assimilated to the voltage bias in the experiment and
$r$ is the distance to the center of the sample, located just
``below'' the STM tip. The spin-spin interaction is antiferromagnetic
($J=-1$) for $0<M<M_0$ and ferromagnetic ($J=2$) for $M>M_0$,
$M_{0}$ is $60\%$--$70\%$ of the maximum possible value of the magnetization. The correlation length $\xi$ may also change
discontinuously and, counterintuitively, the temperature decreases as
the tunneling current increases \cite{schoelz15}. The $M$
  versus $h$ curve of this model is as follows \cite{schoelz15}. At zero field,
$M=0$ and thus $J=-1$. As $h_{0}$ increases to
$h_{0}^{\text{max}}=3$, the spin-spin interaction reverses suddenly to
ferromagnetic ($J=2$) at a field $h_{0}\simeq 2.5$ for which $M$ has
reached $M_{0}$. This discontinuous increase in $J$
at $h_{0}=2.5$ causes a sudden increase of the
magnetization. Afterwards, when the external field is decreased back
to $0$, the coupling constant is left unchanged at $J=2$ and therefore
the spins never go back to the initial state. To further 
mimic experimental results, a smaller jump in the magnetization for
$h_0<2.5$ is induced by an increase in the correlation length $\xi$;
see Figure 3 in ref.~\cite{schoelz15}.

In this paper, we qualitatively explain Schoelz {\em et al}'s
experimental findings \cite{schoelz15} by using a spin-membrane model
that exhibits ripples on a flat membrane, buckling and a dynamical
transition from floppy to rigid states.  Thus we do not need to: (i) interpret spins as many-atom ripples, (ii) introduce jumps
  in $J$ and $\xi$ with $M$, and (iii) decrease the temperature with increasing
  tunneling current, as done in ref.~\cite{schoelz15}. Our
model includes coupling between out-of-plane elastic displacements of atoms and
 local pseudo-spins that pull atoms off plane. The pseudo-spins
are coupled by nearest neighbor interactions. In a previous
publication, we have analyzed a similar model under constant,
  low, temperature conditions \cite{Ruiz-Garcia}. STM experiments
occur under varying temperature conditions because of Joule heating
due to the tunneling current. Increasing the temperature is akin to
driving the system through a first-order phase transition, which is
the essence of our explanation of Schoelz et al's experiments. Thus in
the present work we include: (a) an external field that represents the
STM voltage and, most importantly, (b) the (non-homogeneous)
time-dependent temperature profile brought about by the STM heating of
the sample. For different values of control parameters, first and
second order phase transitions between a rippled-flat membrane state
and a buckled state appear. In the parameter region where these phases
coexist, it is possible to drive the system in conditions that mimic
those in the experiment: inhomogeneous sample heating due to the
tunneling current and electrostatic tip-sample interaction
\cite{schoelz15}. We then show that the wrinkled to buckled transition
appears naturally in our model, without having to invoke ad-hoc jumps
in the model parameters. Moreover, the spin-membrane model
reproduces all the key experimental observations in the STM
experiment, while providing a reasonable physical picture of the real
system.

The plan of the paper is as follows. The spin-membrane model in a
hexagonal lattice is introduced in section~\ref{2d}. The different
equilibrium phases are numerically characterized in
section~\ref{2d-phase-diag}, in which we show that there is a first
order phase transition between a flat but rippled membrane and a
buckled one. In section~\ref{wrinkled-to-buckled}, we drive the system
through the first order phase transition in conditions similar to
those in the experiments. A discussion of our results is presented in
section~\ref{conclusions}.

\section{The model on a hexagonal lattice}\label{2d}

Here, we briefly present our 2d model and its governing
equations. Similar models include simpler 0d spin-oscillator
\cite{jstat10,jstat10a}, 1d spin-string \cite{pre12bon} and 2d
spin-membrane \cite{jsm12bon} models. More complex models include
spin-membrane coupling as well as next neighbor and nearest next
neighbor spin-spin couplings \cite{Ruiz-Garcia}. All these models
exhibit phase transitions between a flat membrane state and a buckled
state below some critical temperature. Additional transitions occur in
the model that has short-ranged spin-spin interactions
\cite{Ruiz-Garcia}. There are different phases characterized by two
order parameters: the magnetization and a domain length parameter that
gives information about the pseudo-spins spatial correlations. For a
hexagonal lattice, there are buckled phases with non-vanishing global
magnetization and also rippled phases with zero magnetization
\cite{Ruiz-Garcia}. The pseudo-spins are partially correlated in space
in these latter phases, which comprise long
wavelength phases, analogous to those in
refs.~\cite{mey07,pss09ban}, stripy phases as in
\cite{ASC11mao}, and atomic wavelength phases, similar to the
ordered phases in ref.~\cite{nl12oha}.  Earlier theoretical studies
of buckling in membranes include the existence of a critical
temperature for buckling in polymerized sheets \cite{Guitter88} and
buckling in graphene due to doping \cite{gazit09}.

\begin{figure}
\centering
\includegraphics[width=3in]{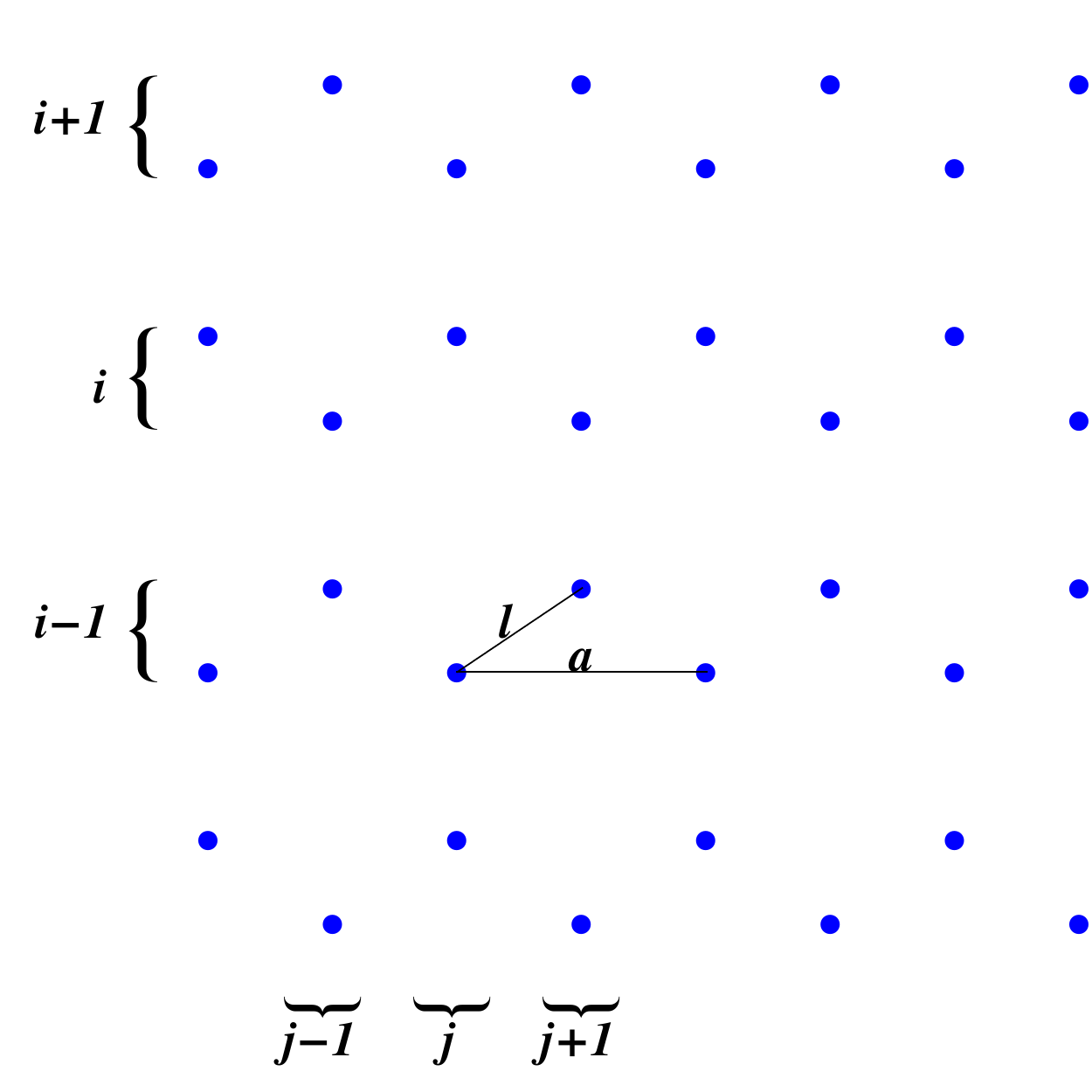}
\caption{Figure summarizing the atom indices and the
  parameters of the unit cell of the hexagonal lattice. In doing this
  plot, we have indexed files and rows assuming that $|i-j|$ is even.}
\label{fig2}
\end{figure}

Carbon atoms are placed on a hexagonal lattice as shown in
Fig.~\ref{fig2}. Let $(\sigma_{ij}$, $u_{ij}$, $p_{ij})$ be the values
of the atom pseudo-spin, height and momentum, respectively, at site
$(i,j)$.  The Hamiltonian is
\begin{align}
  \mathcal{H} =\sum_{ij}
                 \left(\frac{p_{ij}^2}{2m}-fu_{ij}\sigma_{ij}\right)
                 + \sum_{|i-j|=\text{even}} & \Bigg\{ \frac{k}{2}\big[
                 (u_{ij}-u_{i+1,j})^2+
                 (u_{ij}-u_{i,j-1})^2 
                 +(u_{ij}-u_{i,j+1})^2  \big] \nonumber \\
               & + J                 \sigma_{ij}(\sigma_{i+1,j}+\sigma_{i,j-1}+\sigma_{i,j+1})\Bigg\}. 
\label{H}
\end{align}
This is a particular case of the Hamiltonian introduced in
ref.~\cite{Ruiz-Garcia} that had an additional
next-nearest-neighbor interaction among spins.

The dynamics of the system consists of
(i) Hamilton's equations of motion for $(u_{ij},p_{ij})$, and (ii) Glauber dynamics \cite{Gl63} at
temperature $T$ for $\sigma_{ij}$: 
\begin{align}
 &\ddot{u}_{ij} - K_N^2 (u_{i+1,j}+u_{i,j-1}+u_{i,j+1}-3u_{ij})= \sigma_{ij}, \label{mov}\\
&\omega_{ij}(\bm{\sigma}|\bm{u})=\frac{\delta}{2}(1-\gamma_{ij}
\sigma_{ij}), \label{omega}\\
& \gamma_{ij}=\tanh\left[\frac{u_{ij}}{\theta} -\frac{\kappa}{\theta}(\sigma_{i+1,j}+\sigma_{i,j+1}+\sigma_{i,j-1})\right]\!.
\end{align}
Here $\omega_{ij}$ is the rate at which the pseudo-spin at site
$(i,j)$ flips and $\delta$ is a parameter setting the characteristic
time-scale for the pseudo-spin flips. In the long time
limit, the system reaches thermodynamic equilibrium and its
probability distribution has the canonical form
$\mathcal{P}\propto \exp(-\mathcal{H}/T)$. It is convenient to
introduce the following parameters,
\begin{equation}
T_0 =\frac{f^2K_{N}^{2}}{k}, \quad K_N =\frac{3n-2}{\sqrt{6}\pi},
\end{equation}
where $n$ is the total number of rows in the lattice. The temperature
$T_{0}$ is the transition temperature from a (high temperature) flat to a buckled string
configuration for $J=0$ \cite{Ruiz-Garcia}. Then, we define
dimensionless displacements and time,
\begin{equation}
  u_{ij}^*=\frac{k u_{ij}}{f K_{N}^2},\qquad
  t^*=\frac{t}{K_{N}}\sqrt{\frac{k}{m}},\label{nondim_u_t}
\end{equation}
and also dimensionless spin-spin coupling constant and temperature,
\begin{equation}\label{kappa-theta}
\kappa=\frac{J}{ T_{0}}, \quad \theta=\frac{T}{T_{0}}=\frac{k\,T}{f^{2}K_{N}^{2}}.
\end{equation}
Thus we measure energy in units of the transition temperature $T_{0}$.

In the equilibrium state, the average dimensionless displacements 
obey the discrete Poisson equation
\begin{equation}
 - K_N^2 (u_{i+1,j}+u_{i,j-1}+u_{i,j+1}-3u_{ij})= \mu_{ij}, \label{mov_av}
\end{equation}
in which $\mu_{ij}$ stands for the average magnetization at site
$(i,j)$. The asterisks have been omitted so as not to clutter the
formulae. In the continuum limit, Eq.~\eqref{mov_av} becomes
\begin{equation}\label{curvature}
  \frac{1}{2\pi^{2}}\nabla^{2}u(x,y)=\mu(x,y).
\end{equation}
Here $0\leq x,y\leq 1$, and the sample becomes the unit square in the
continuum limit with our choice of dimensionless variables
\cite{Ruiz-Garcia}. Therefore, the average magnetization gives the
curvature of the membrane. Thus we can deduce the state of the membrane by looking at either the atoms displacements $u(x,y)$ or the pseudo-spins local value $\mu(x,y)$.

\section{Equilibrium Phase Diagrams}\label{2d-phase-diag}

\begin{figure}
\centering
\subfigure[~Magnetization]{\includegraphics[width=3in]{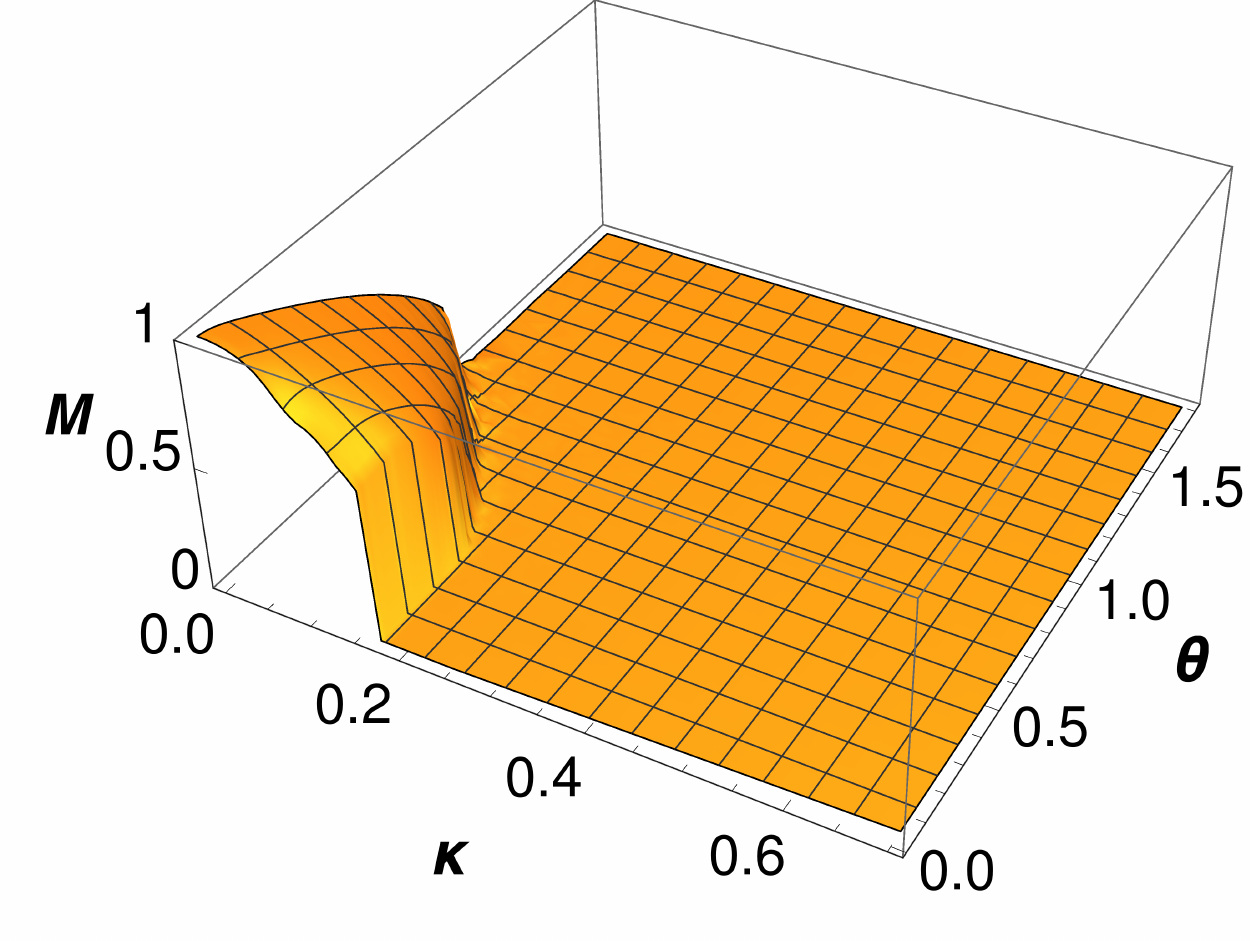}}\hspace{2em}
\subfigure[~Domain Length]{\includegraphics[width=3in]{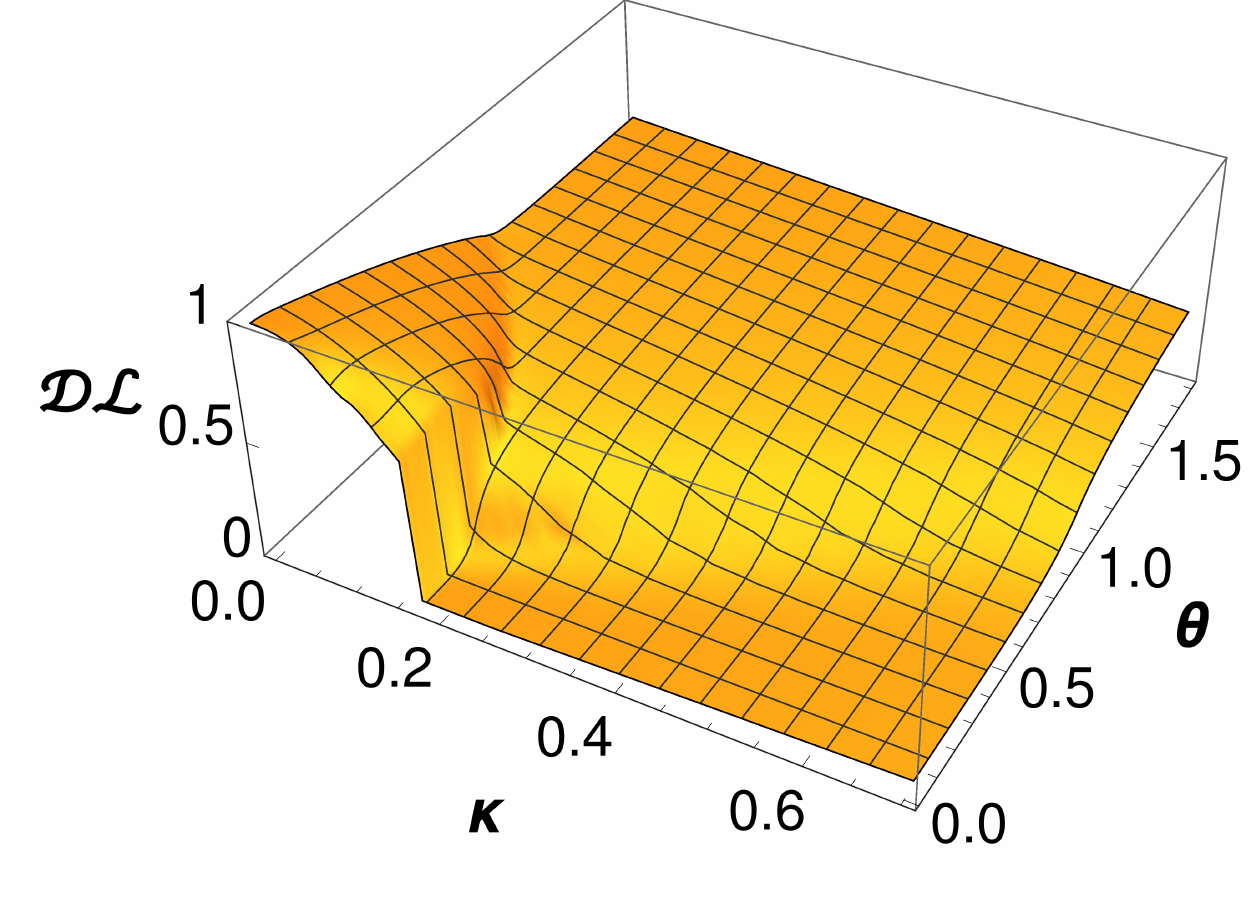}}
\caption{(a) Magnetization and (b) domain length parameter as functions of $\theta$ and $\kappa$ obtained by temperature downsweeping as explained in the text.} 
\label{fig_fd_1}
\end{figure}

Except for $J=0$ that can be exactly solved, the equilibrium phase
diagrams have to be calculated numerically. At $J=0$, the flat
solution bifurcates at $T=T_{0}$ to a buckled state, which is
thermodynamically stable for $T<T_{0}$ \cite{Ruiz-Garcia}. This can be
appreciated in Fig.~\ref{fig_fd_1}, which has been drawn by
down-sweeping the dimensionless temperature from a given $\theta>1$ at
each fixed value of $\kappa$. At the largest value of $\theta$, the
initial configuration is random and the simulation reaches equilibrium
after a certain time. Then, the magnetization $M$ and the domain length
parameter $\mathcal{DL}$ of ref.~\cite{nl12oha},
\begin{eqnarray}\label{orderparam}
  M&=&\left|\frac{1}{N}\sum_{ij}\sigma_{ij}\right|\!, \\
  \mathcal{DL}&=&\frac{1}{3N}\sum_{|i-j|=\text{even}}
                  \left[3+\sigma_{ij}\left(\sigma_{i+1,j}+\sigma_{i,j-1}+\sigma_{i,j+1}
\right)\right].
\end{eqnarray}
are registered. For the next simulation, $\theta$ is slightly lowered
and the equilibrium configuration reached at the previous temperature
is used as the initial condition. This procedure is continued until
the phase diagram is completed. The parameter $\mathcal{DL}$ gives
information about the difference between the number of ferromagnetic
(contributing +1 to $\mathcal{DL}$) and antiferromagnetic
(contributing -1 to $\mathcal{DL}$) links and makes it possible to
discriminate between different phases with zero global
magnetization. Specifically, we have $\mathcal{DL}=1/2$ for random
pseudo-spins and $\mathcal{DL}=0$ for antiferromagnetic ordering. For
ferromagnetic ordering, it is $M=\mathcal{DL}=1$. \textrm{Note that
  the magnetization \eqref{orderparam} does not discriminate between
  the two possible signs of the curvature in Eq.~\eqref{curvature}. }

\begin{figure}
\centering
\subfigure[~Magnetization]{\includegraphics[width=3in]{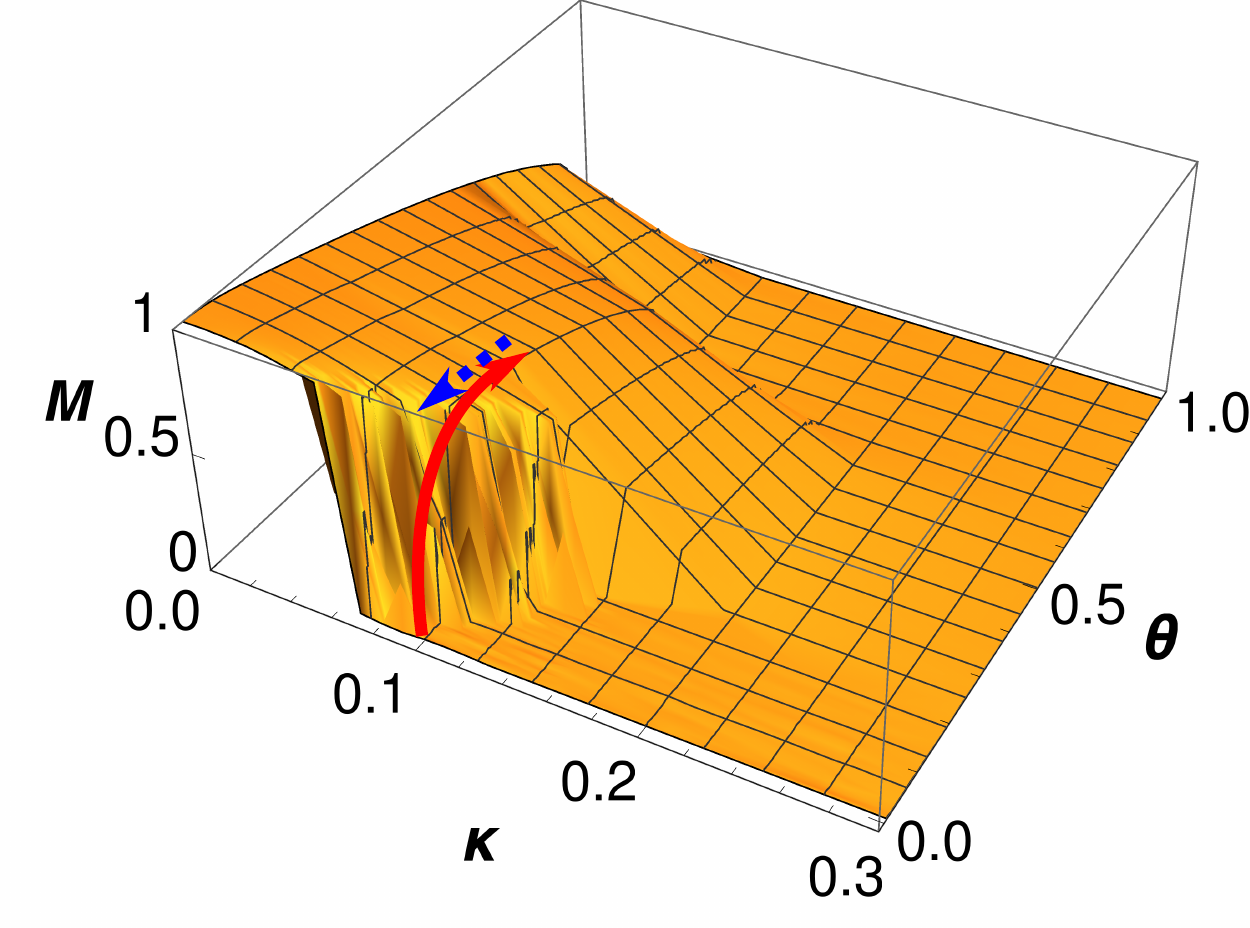}}\hspace{2em}
\subfigure[~Domain Length]{\includegraphics[width=3in]{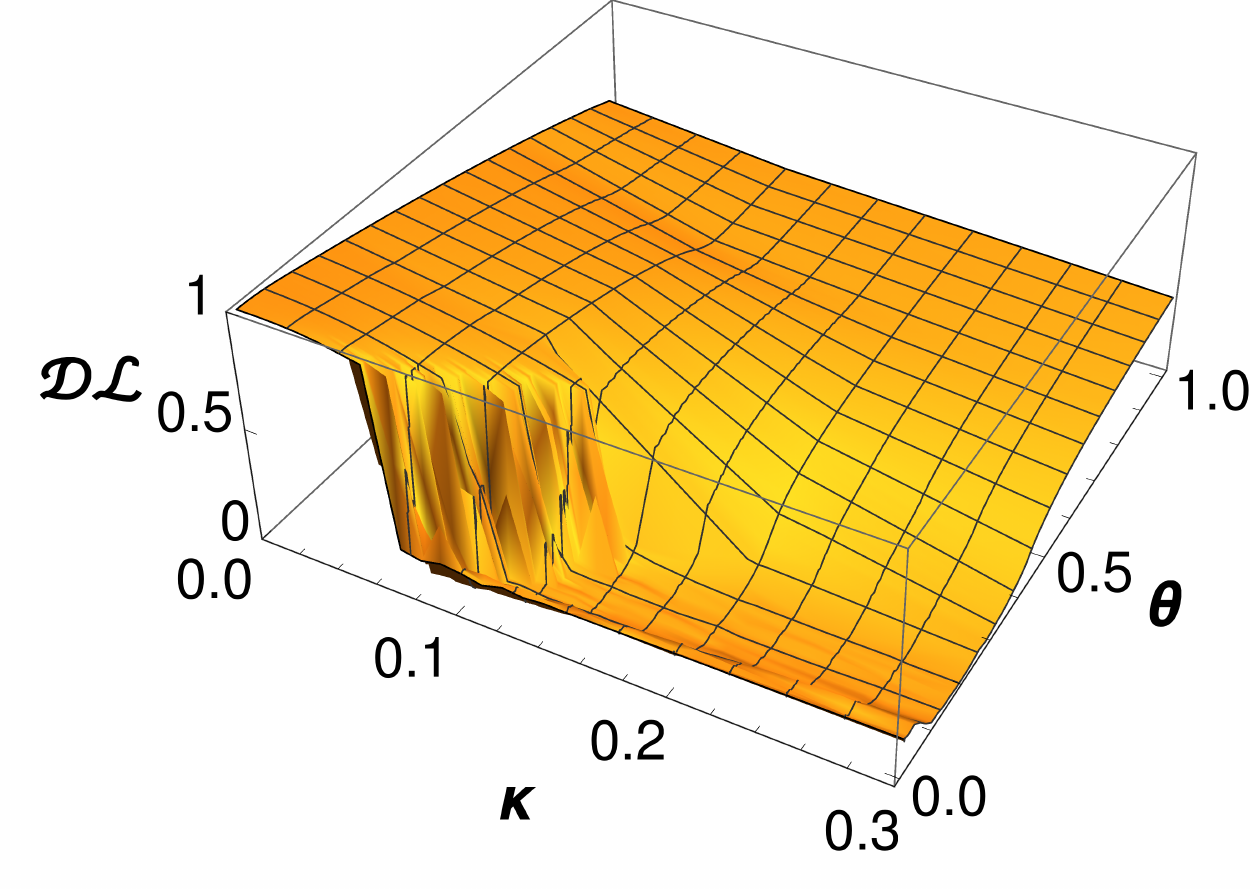}}
\caption{(a) Magnetization and (b) domain length
  parameter as functions of $\theta$ and $\kappa$. The
  initial configuration for all the simulations consists of a flat
  membrane and randomly oriented pseudo-spins. In (a), we show a
  heating cycle corresponding to \textrm{the} interaction with a STM
  tip\textrm{, see Sec.~\ref{wrinkled-to-buckled}. T}he {continuous }red arrow represents the part of the cycle where
  \textrm{the} temperature is increased and the \textrm{wrinkled to
    buckled} phase transition occurs, \textrm{whereas} the {dashed} blue arrow
  \textrm{marks} the cooling part of the cycle where no transition is
  found \textrm{since the system remains buckled}.}
\label{fig_fd_2}
\end{figure}

The method we have just described produces the correct phase diagram
provided the phase transitions are second order, which is the case for
high critical temperatures ($\theta>0.5$ as seen in
Fig.~\ref{fig_fd_1}). For first order phase transitions, down-sweeping
yields only one part of the hysteresis loops associated with first
order phase transitions, specifically that corresponding to the stable
phase at the higher temperatures. To visualize the thermodynamically
stable phase at first order phase transitions that occur for low
critical temperatures, $\theta<0.5$, we have redrawn the diagram
always starting simulations from a random configuration and waiting
for the system to equilibrate. This produces Fig.~\ref{fig_fd_2}. On
the one hand, we observe that there is a region of zero magnetization
at low temperatures (approximately, $0.07<\kappa<0.2$) that was absent
in Fig.~\ref{fig_fd_1}. In this region, the membrane is rippled as
shown by its partial antiferromagnetic ordering,
$0\leq{\cal DL}\leq 0.1$.  On the other hand, the membrane ends up in
low temperatures states that are similar to those in
Fig.~\ref{fig_fd_1} both for $\kappa\lesssim 0.07$ (buckled membrane)
and $\kappa\gtrsim 0.2$ (rippled flat membrane) \footnote{In
  ref.~\cite{Ruiz-Garcia}, we study a related model that has an
  extra nearest neighbor spin-spin coupling $\lambda$. We calculate numerically phase diagrams as a function of $\kappa$ and
    $\lambda$ for a constant low temperature value
    $\theta=0.01$.}.

\section{Driving graphene across the rippled to buckled phase transition}\label{wrinkled-to-buckled}

In Schoelz {\em et al}'s experiments \cite{schoelz15}, the floppy
rippled membrane undergoes a transition to a rigid buckled state when
heated by the STM current. In our model, this may correspond to
driving the system across the low temperature first order phase
transition seen in Fig.~\ref{fig_fd_2} for small values of $\kappa$
and $\theta$. To illustrate this, we set $\kappa=0.1$ and
$\theta= 0.01$ for all lattice points in our numerical simulations and
start with an initially flat membrane and randomly oriented
pseudo-spins. We consider $n=35$ rows in a 2d hexagonal lattice, with
$N=2100$ atoms. The system reaches a stationary state, which is
typically rippled, $M=0$ and ${\cal DL}\simeq 0.1 <1/2$.

First, for the sake of simplicity and to understand the basic physical
mechanism under the first order transition, we analyze \textit{homogeneous heating} of the membrane. Second, in order to have a situation closer to the experiments and discuss some more specific details thereof, we consider the case of \textit{inhomogeneous heating}.

\subsection{Homogeneous heating}\label{homogeneous-heating}

Assume that the heat bath temperature felt by the pseudo-spins is the
same at all lattice points and varied at a constant rate.  The
pseudo-spins flip according to the Glauber dynamics given in
\eqref{omega} with the instantaneous and externally controlled value
of the temperature $\theta(t)$.

\begin{figure}
\centering
\subfigure[~Magnetization]{\includegraphics[width=0.45\textwidth]{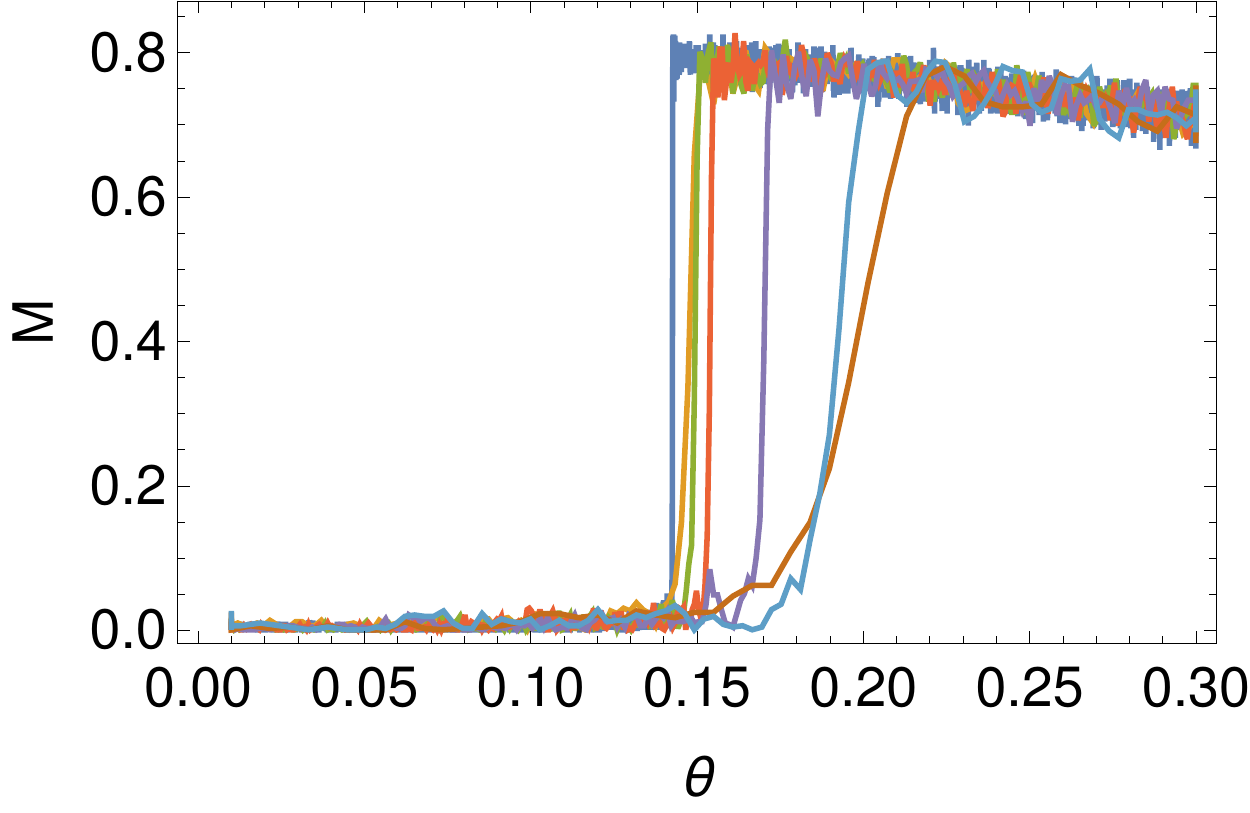}}
\subfigure[~Central Atom Height]{\includegraphics[width=0.45\textwidth]{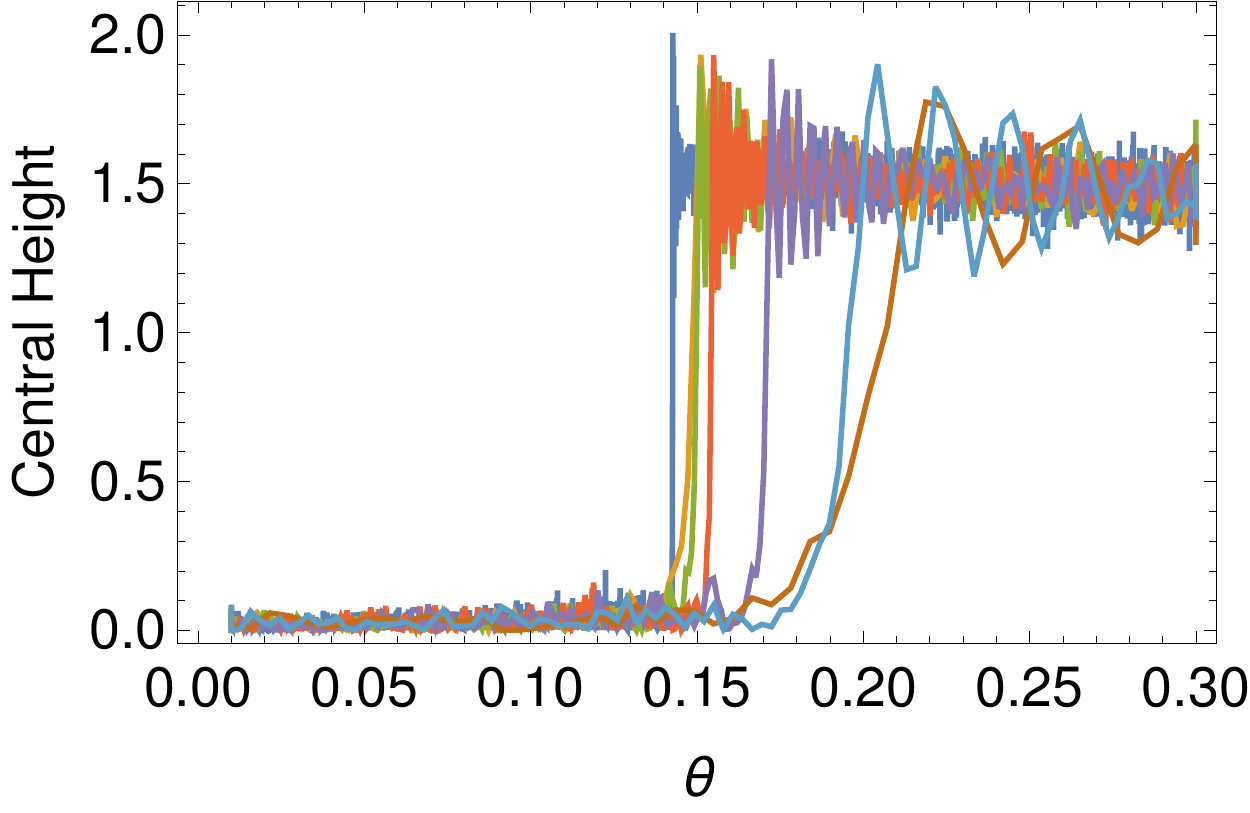}}
\caption{Magnetization and height of the central atom as a function of
  temperature, upon heating of the system.  Different lines
  correspond to different heating rates. Both the magnetization and
  the central atom height jump around $\theta_{2}=0.15$,
  revealing a first-order phase transition in the system. The slower
  the system is heated, the lower the temperature of the jump. In
  these simulations the initial and final temperatures are
  $\theta_{0}=0.01$ and $\theta_{f}=0.3$, respectively. The system is
  heated with a constant rate, $\theta(t)=\theta_{0}+rt$, and the
  different lines correspond to rates (from left to right):
  $3\times 10^{-5}$, $3\times 10^{-4}$, $4\times 10^{-4}$,
  $6\times 10^{-4}$, $10^{-3}$, $3\times 10^{-3}$, $6\times 10^{-3}$.}
  \label{fig_ph_tr}
\end{figure}

Upon heating, the membrane remains rippled with zero magnetization for
$\theta\lesssim 0.15$. At about $\theta_{2}=0.15$, the magnetization
and the height of the central atom suddenly increase, as shown in
Fig.~\ref{fig_ph_tr}. This effect strongly resembles the STM
experiments in ref.~\cite{schoelz15}, where the increase in
dissipated power (modeled here with an increase of the temperature of
the heat bath to which the system is coupled) promotes a discrete
increase in height, that is, a buckling transition.

The temperature at which the transition occurs depends on the heating
rate. For the slowest rates, the jump is almost vertical and takes
place at $\theta\simeq 0.15$. For faster rates, the transition is
softer and happens for a slightly higher temperature, up to
$\theta\simeq 0.20$ for the values considered in
Fig.~\ref{fig_ph_tr}. The physical image is the following: a very
slow, almost quasi-static, process leads to a sharp transition at the
temperature at which the flat membrane becomes unstable. If heating is
faster, the system remains in the unstable flat configuration for a
certain time and is hindered from finding the ``path'' to the true
thermodynamic equilibrium.

\begin{figure}
\centering
\includegraphics[width=3.25in]{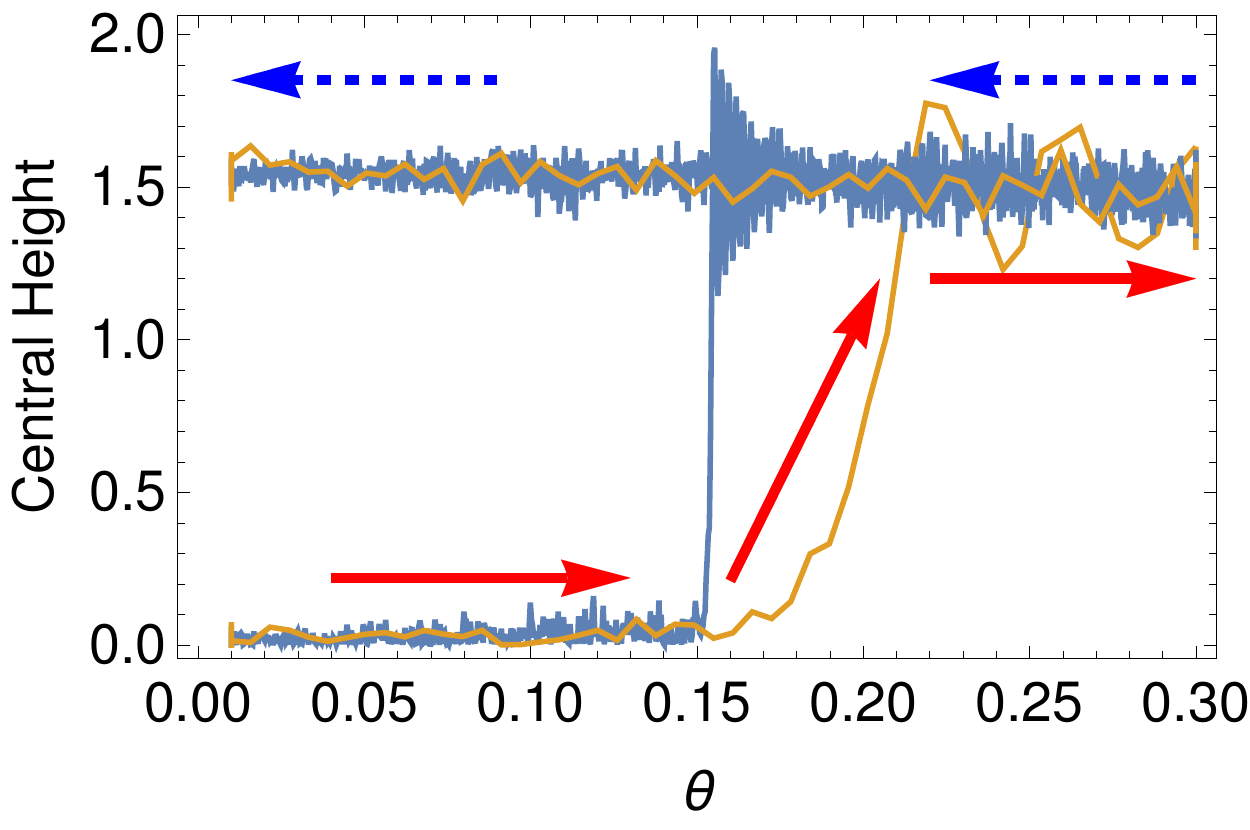}
\caption{Height of the central atom as a function of the temperature
  for two heating/cooling cycles, {the leftmost and rightmost lines
    correspond with }the slowest and fastest rates of temperature
  variation in Fig.~\ref{fig_ph_tr}, respectively. {The
    solid arrows and the dashed ones mark the
    heating and cooling part of the cycle respectively, analogously to
    the arrows present in Fig.~\ref{fig_fd_2}. } When cooled, the
  system remains buckled for $\theta<0.15$, which shows that the flat
  rippled membrane configuration in the low temperature region is
  metastable.  }
\label{fig_ph_tr_2}
\end{figure}

Finally, in Fig.~\ref{fig_ph_tr_2}, we present simulations in which
the temperature is first increased, until the membrane buckles, and
the system is subsequently cooled down to the initial low
temperature. Interestingly, we observe that the system remains buckled
when the temperature is lowered. This hysteretic behavior is a
numerical proof of the metastability of the initial wrinkled
configuration for low temperatures and thus is consistent with
Fig.~\ref{fig_fd_1}.  The final state resembles the ``rigid'' states
that are reached in STM experiments for large enough currents
\cite{schoelz15}.

\subsection{Inhomogeneous heating}\label{inhomogeneous-heating}

In STM experiments, the graphene sample is \textit{locally} heated. We
model this by an inhomogeneous temperature profile of a circular
membrane of radius $R=1/2$ (clamped at the boundary) inscribed in the
unit square.  Throughout this section, $\bm{r}$ stands for any point
in the circle, $0\leq r\leq R$, with $r=|\bm{r}|$. Energy is injected
at the membrane center and the temperature is initially homogeneous
throughout the sample, $\theta(\bm{r},t=0)=\theta_{0}$.  At $t=0$, the
heating process starts, and the border of the sample is always kept at
room temperature $\theta_{0}$, $\theta(\bm{r},t)|_{r=1/2}=\theta_{0}$.

The space and time temperature profile obeys the heat equation with a
source term,
\begin{equation}
\partial_{t}\theta- \alpha \nabla^2 \theta = q(\bm{r}), \quad q(\bm{r})=Q_{0} \, e^{-r^{2}/a^{2}},
\label{q-expression}
\end{equation}
Note that we are using dimensionless variables, so that the thermal
diffusivity $\alpha$ and the energy source from the STM tip $q$ are
measured in the units introduced in the previous sections (the
dimensions of $\alpha$ and $q$ are $\text{length}^{2}/\text{time}$ and
$\text{energy}/\text{time}$, respectively). The source term has radial
symmetry and exhibits a Gaussian decay from its maximum value $Q_{0}$
over a characteristic length $a$ (in dimensional units, $a$ is a few
angstroms \cite{Chen-book}). Note that, for fixed values of $a$ and
$\alpha$, the total injected power is proportional to
$Q_{0}$. Therefore, we can consider that $Q_{0}\propto IV$ in the STM
experiments, where $I$ is the tunneling current and $V$ the voltage
bias between the tip and the sample.  Interestingly, the same lateral
decay of the injected power has been used in other experimental
situations, see for instance ref.~\cite{acs10fau} for the study
of the thermal conductivity of a graphene membrane excited by a laser.

\begin{figure}
\centering
\includegraphics[width=3.25in]{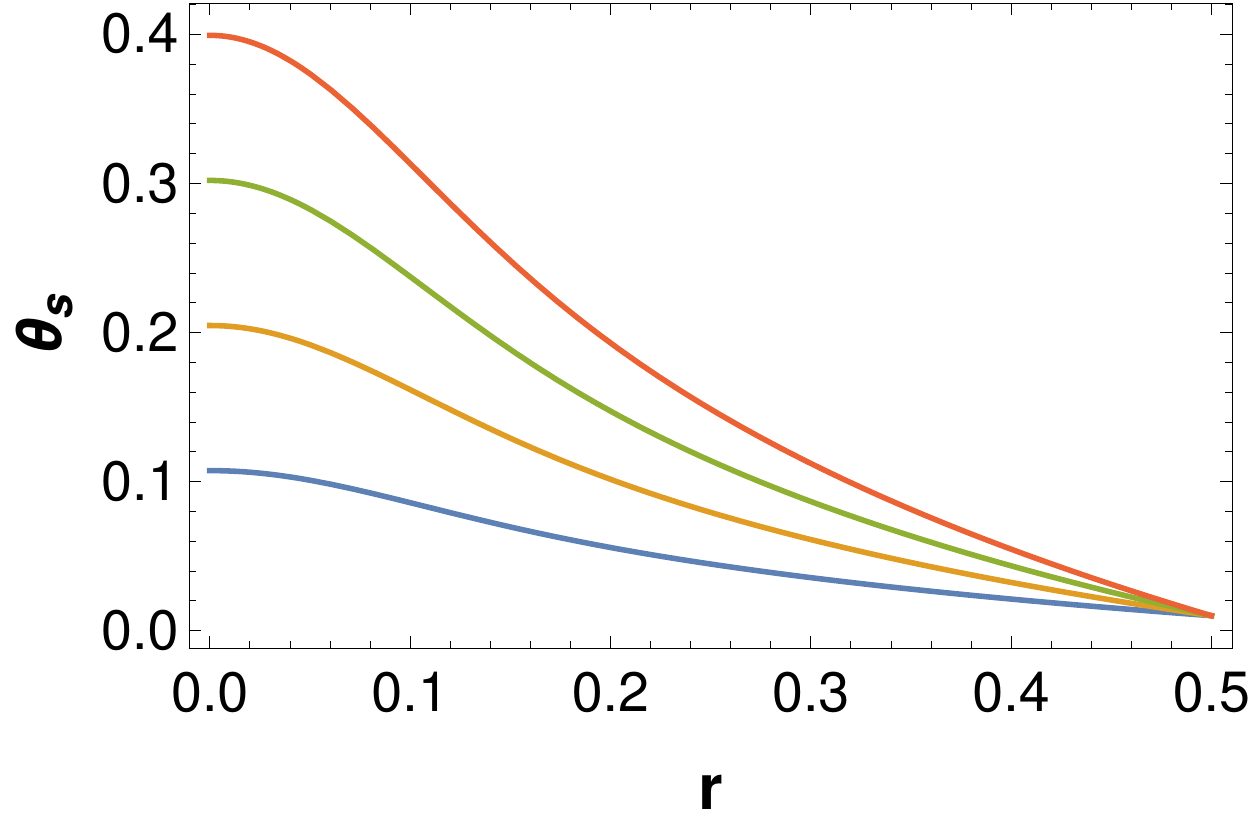}
\caption{From bottom to top, steady temperature profiles for
  $\Delta\theta=0.1$, $0.2$, $0.3$, $0.4$.  }
\label{fig:temps}
\end{figure}

We seek stationary solutions of the heat equation with radial
symmetry, $\theta(r,t)=\theta_s(r)$,
which obey
\begin{subequations}\label{seq}
\begin{equation}
\nabla^{2}\theta_s+\frac{Q_{0}}{\alpha}\,
e^{-r^2/a^{2}}=0, \\
\end{equation}
\begin{equation}
\theta_s(r=R)=\theta_{0}, \quad \lim_{r\to 0}|\partial_{r}\theta_{s}(r)|<\infty.
\end{equation}
\end{subequations}
Equation~\eqref{seq} is solved along the same lines as in ref.~\cite{acs10fau}, with the result
\begin{equation}
  \theta_s(r)=\theta_{0}+\frac{\Delta\theta}{2}\int_{r/a}^{R/a} dx\,
  \frac{1-e^{-x^2}}{x}, \quad \Delta\theta=\frac{Q_{0} a^{2}}{\alpha}.\label{T_s}
\end{equation}
We plot this stationary temperature profile for several values of
$\Delta\theta$ in Fig.~\ref{fig:temps}.  We do not consider the
transitory decay of the temperature profile to this steady solution,
since graphene is a very good thermal conductor \cite{acs10fau}. Thus,
we expect the time scale for the decay to this steady profile to be
much shorter than those associated to the increase or the decrease of
the voltage bias in the STM experiments.  In any case, we would like
to stress that taking into account the transient to the stationary
state does not alter our conclusions.

The STM tip also has an electrostatic interaction with the sample,
which is included in our model by adding an external-field term
$\Delta{\cal H}$ to the Hamiltonian \eqref{H},
\begin{equation}\label{delta-H}
 \Delta{\cal H}=-\sum_{ij}h_{ij}\sigma_{ij}.
\end{equation}
\textrm{Note that the external field breaks the up-down symmetry of
  the pseudospins, which gives rise to a preferred sign of the
  curvature in Eq.~\eqref{curvature}.} In ref.~\cite{schoelz15}, the
field $h_{ij}$ decays exponentially from the center of the tip over a
characteristic length of a few hundreds of the graphene lattice
constant, which is consistent with the long-range character of the
electrostatic interaction. In our work, we consider samples with
$1650$ atoms inside the circle of unit diameter. For such \textrm{small}
samples, the field experiences almost no decay and, therefore, we
simply take $h_{ij}=h$, independent of $(i,j)$. Since the strength of
the electrostatic interaction increases with the applied bias $V$, we
identify $h$ with $V$. Thus, the current is $I=\Delta\theta/V$ and the
width of the source term, $a$, is three lattice constants in our
simulations. \textrm{For the sake of clarity, we sum up the key
  parameters of the model that control the behaviour showed in the
  simulations in Table~\ref{tab_1}. }

\begin{table}[]

\centering

\caption{{Main parameters controlling the behavior of the system
    in the inhomogeneous heating process, for an applied bias $V$ and
    a tunneling current $I$ spread over a region of characteristic
    length $a$, as expressed by Eqs.~\eqref{q-expression}-\eqref{delta-H}.}}

\label{tab_1}

\begin{tabular}{| l | p{2.5in} | p{2.5in}|}
  \hline
Parameter & \protect{\centering{Role}} & \protect{\centering{Controlling}}
  \\
\hline
  $\kappa$ & Pseudospins' antiferro
             interaction  & Lower critical
             temperature
             $\theta_{2}$,
             the
             system buckles for $\theta>\theta_{2}$.\\
  \hline
$\Delta \theta\propto I V a^{2}$ & Strength of the Joule effect & 
                                   Temperature at  the
                                   center of the sample: should be
                                                                  larger
                                                                  than 
                                                                 $\theta_{2}$
                                                                  to
                                                                  induce
                                                                  buckling.\\
                                                            \hline
$h$ & Strength of the tip-sample electrostatic interaction
                          &   Sign of the curvature (breaks up-down symmetry). \\                     
\hline
\end{tabular}

\end{table}

\begin{figure}
\centering
\includegraphics[width=3.25in]{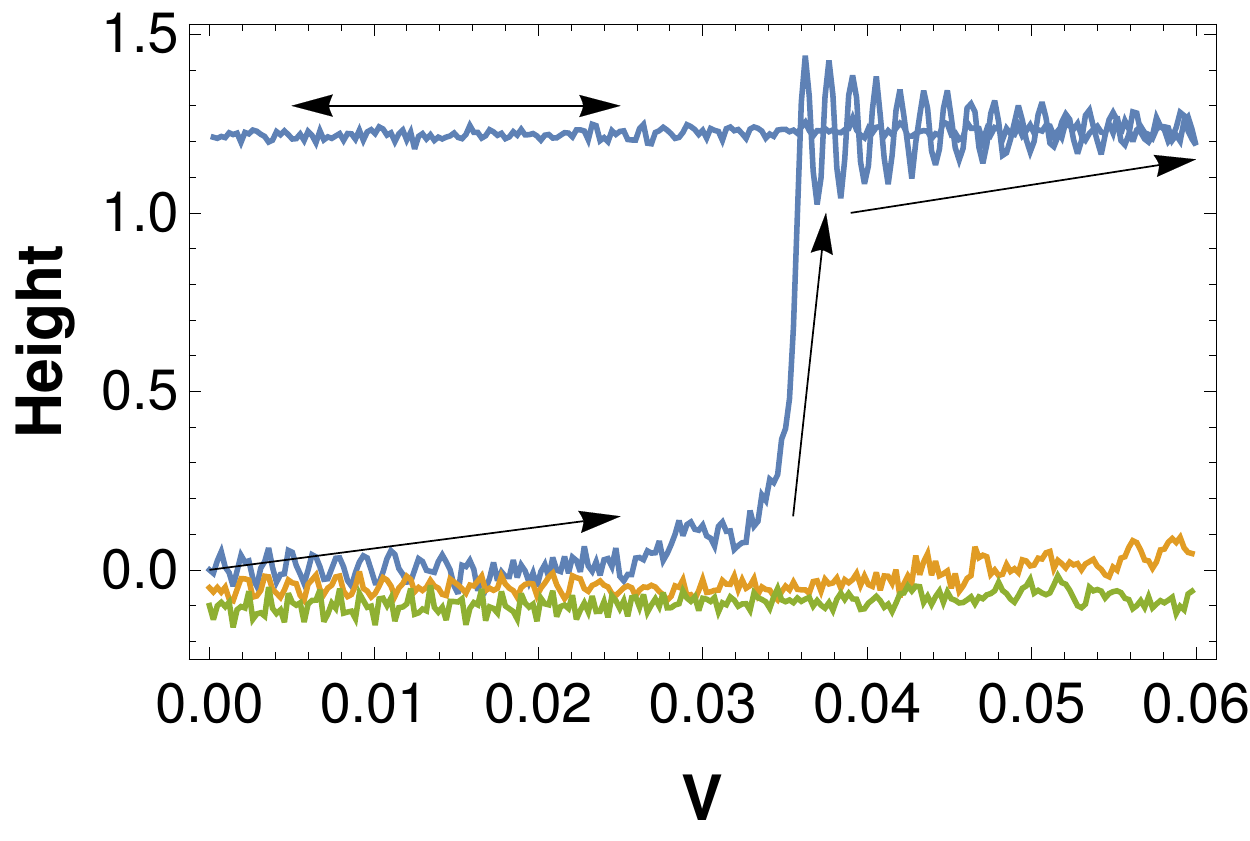}
\caption{Central atom height vs.~voltage bias $V$. The simulations
  have been conducted in a circular membrane having $N=1650$
  sites. From top to bottom, the lines correspond to tunneling
  currents $I=\Delta\theta/V=1$, $2$, and $6$.  The voltage $V$
  increases at a constant rate from $0$ to $V_{\max}$ in steps
  $V_{\max}/250$, during a total time $t_{total}=250$. For clarity, we
  have shifted downwards the two lowest curves.  The curve for $I=6$
  exhibits a small bump around $V\simeq 0.025$, which coincides with
  the maximum of $\theta_s$ reaching $0.15$. Once we have reached
  $V_{\max}$ for $I=6$, we decrease back the voltage to 0 and then
  height follows the almost flat upper curve. This shows that the jump
  at $V\simeq 0.035$ is irreversible.}
\label{I-V}
\end{figure}

To mimic the experimental procedure in ref.~\cite{schoelz15}, we
fix $I$ in each simulation, increase $V$ at a certain constant rate
and track the height of the central atom, see Fig.~\ref{I-V}. In this
way, we are driving the system in the parameter region where there is
a first order phase transition as described in the previous
section. Except for not having averaged the oscillations in our
numerical results, the behavior displayed in Fig.~\ref{I-V} is
completely analogous to that observed in ref.~\cite{schoelz15},
see Fig.~\ref{schoelz}. For small $I$, the increase in $V$
produces a reversible pulling that increases the global magnetization
and the height of the central atom but does not produce overall
buckling. Here, reversible means that if the voltage is decreased back
to zero from its maximum value, the same curves are swept. This
notwithstanding, once $I$ reaches a certain critical value,
\textit{non-reversible} buckling appears (upper-curve): the membrane
remains buckled when the voltage is decreased back to zero.

In the STM experiments, the buckling (when it occurs) comprises two
steps: apart from the large jump in height at a certain value of the
voltage $V_{c}$, there appears a smaller ``bump'' in height at a
smaller voltage $V_{1}<V_{c}$. Interestingly, even this fine detail of
the experimental results is reproduced by our model without having to
assume a jump in the correlation length $\xi$ as in
ref.~\cite{schoelz15}. As energy is injected, first the maximum
of the temperature profile (at the center $r=0$) exceeds the critical
value $\theta_{2}\simeq 0.15$ at $V\simeq V_{1}\simeq 0.025$ and this
brings about the small height bump observed in Fig.~\ref{I-V} between
$V=0.025$ and $V$=0.035. Second, as the voltage bias is further
increased to $V_{c}>V_{1}$, there is a large enough region of the
system in which the temperature is above $\theta_{2}$, which makes the
system buckle.

In the considered range of $V$, $0\leq V\leq 0.06$, heating
($I\neq 0$) is absolutely necessary to produce membrane buckling
because the external field is not strong enough by itself. However, if
we further raise $V$, it would reach a value at which the system
buckles even without heating ($I=0$). Therefore, our model may also be
useful to investigate the buckling phenomena observed when strong
electrostatic forces are applied, as in
refs.~\cite{lin12,svensson11}.

It is worth stressing some further aspects of our numerical
  results in Fig.~\ref{I-V}. First, we increase the voltage at a
  specific rate and, therefore, different curves are obtained for
  different rates. Of course, a rate-independent equilibrium curve is
  obtained if the voltage is increased slowly enough, that is,
  quasi-statically. Second, our numerical results show some time
  oscillations. Therefore, the present model allows us to resolve the
  time evolution of the membrane over a finer scale than that of the
  currently available experimental results, which are time-averaged.

\section{Discussion}\label{conclusions}

Our spin-membrane model exhibits a first order phase transition from
rippled-flat to buckled membrane for appropriately small values of the
non-dimensional temperature and spin-spin coupling. The main parameter
to be fixed is $\kappa$, that is, the strength of the
antiferromagnetic pseudo-spin interaction. Once $\kappa$ has been
chosen in the range where the low-temperature first order
phase transition is present, it also determines the
  temperature $\theta_{2}$ above which the membrane buckles. The
  additional quantities controlling the system behavior are $I$ and
$V$, which govern the strength of the Joule effect that heats
  the membrane, so that the temperature $\theta>\theta_{2}$, and makes
  it buckle. Conversely, the characteristic length $a$ (which
estimates the radius of interaction between the STM tip and the
sample) does not play a key role: changing its value only shifts the
range of $V$ and $I$ over which driving through the transition is
observed.

Membrane buckling arises from the long range interaction
among spins induced by the spin-membrane coupling and the metastable
state of a flat membrane with ripples stems from the short-range
antiferromagnetic spin-spin coupling. To model the results of Schoelz
{\em et al}'s experiments, we need to drive the system through the
first order phase transition by an appropriate control of temperature
and the electrostatic interaction between the STM tip and the graphene
membrane.

Control of a homogeneous bath temperature induces irreversible
buckling but the connection between the parameters of this process and
those in the STM experiment is not transparent. Moreover, the STM
should heat inhomogeneously the sample. Therefore, we have assumed
that the bath temperature adopts the inhomogeneous profile that solves
the heat equation with a Gaussian source term. Furthermore, we have
introduced an external field term in the spin energy that mimics STM
electrostatic force. The latter breaks the spin up-down symmetry
which, in turn (via the spin-membrane coupling), breaks the up-down
symmetry of the vertical membrane displacements.

The combination of the two above mechanisms produces numerical results
that contain every feature of STM buckling experiments, including the
existence of a critical value of the current. Our numerical results
strongly suggest that both the electrostatic force and heat
dissipation are playing a role in the buckling phenomenon observed in
ref.~\cite{schoelz15}. In addition, our spin-membrane model
improves that in ref.~\cite{schoelz15} because it explicitly
shows the membrane ripples and it does not need to change the sign of
the spin-spin coupling to induce buckling.

There are some hurdles that need to be overcome before finding a
microscopic model closer to first principles that explains STM induced
buckling of graphene membranes. Firstly, as experiments become more
accurate, they may allow for a better definition of all parameters in
mesoscopic models, improving the current physical understanding of
graphene rippling. Secondly, starting from an electron-phonon
Hamiltonian for a suspended graphene sheet, it is possible to derive
stationary saddle-point equations for vertical displacements coupled
to some auxiliary fields \cite{gui14}. From these equations, critical
temperatures below which there is buckling can be found
\cite{ByR16}. These results are qualitatively similar to those found
with our spin-membrane model. It seems worth investigating modeling
the interaction between the graphene membrane and the STM tip at the
level of saddle-point equations. Then some inhomogeneous heating
program similar to that in the present paper could be used to explain
Schoelz {\em et al}'s experiments from ``first principles''.

Finally, note that the buckling transition has been also observed in
experiments in which only an electrostatic force is applied to the
sample, with no energy injection. Our model can also explain this
effect, since the external field term favors that the spins have a
well-defined sign, that is, that the sign of the membrane curvature is
well-defined. In this respect, a detailed experimental study of
buckling in graphene, in which both the temperature (via an energy
injection mechanism) and the electrostatic force can be independently
changed, would greatly improve our insight into the internal
interactions that govern buckling.

\acknowledgments
This work has been supported by the Spanish Mi\-nisterio de
Econom\'\i a y Competitividad grants MTM2014-56948-C2-2-P (MRG \& LLB) and FIS2014-53808-P (AP). MRG also acknowledges support from MECD through the FPU program and from MINECO along with Residencia de Estudiantes. AP thanks C.~A.~Plata for
helpful discussions and a careful reading of the manuscript.

\end{document}